\documentclass[11pt,a4paper,twoside]{JHEP3}

\usepackage{cite}
\usepackage{graphicx}
\usepackage{epsfig}
\usepackage{dcolumn}
\usepackage{bm}

\def \d {{\rm d}}

\def \tr {{\tilde\rho}}
\def \tv {{\tilde v}}
\def \tz {{\tilde z}}
\def \e {{\epsilon}}

\newcommand{\be}{\begin{equation}}
\newcommand{\ee}{\end{equation}}

\newcommand{\beqn}{\begin{eqnarray}}
\newcommand{\eeqn}{\end{eqnarray}}

\newcommand{\AS}{Aichelburg-Sexl }
\newcommand{\pa}{\partial}
\newcommand{\pp}{{\it pp\,}-}
\newcommand{\ba}{\begin{array}}
\newcommand{\ea}{\end{array}}

\title{Higher dimensional black holes in external magnetic fields}

\author{Marcello Ortaggio \\ 
        Institute of Theoretical Physics, Faculty of Mathematics and Physics, \\ Charles University in Prague,
  V Hole\v{s}ovi\v{c}k\'{a}ch 2, 180 00 Prague 8, Czech Republic, and \\ 
        INFN (Rome) \\
        E-mail: \email{marcello.ortaggio@comune.re.it}}

\abstract{
We apply a Harrison transformation to higher dimensional asymptotically flat black hole solutions, which puts them into an external magnetic field. First, we magnetize the Schwarzschild-Tangherlini metric in arbitrary spacetime dimension ${n\ge4}$. The thus generated exact solution of the Einstein-Maxwell equations describes a static black hole immersed in a Melvin ``fluxbrane'', and generalizes previous results by Ernst for the case ${n=4}$. The magnetic field deforms the shape of the event horizon, but the total area (as a function of the mass) and the thermodynamics remain unaffected. The amount of flux through a one-dimensional loop on the horizon exhibits a maximum for a finite value of the magnetic field strength, and decreases for larger values. In the Aichelburg-Sexl ultrarelativistic limit, the magnetized black hole becomes an impulsive gravitational wave propagating in the Melvin background. Furthermore, we discuss possible applications of a similar Harrison transformation to rotating black objects. This enables us to magnetize the Myers-Perry hole and the (dipole) Emparan-Reall ring at least in the special case when the vector potential is parallel to a nonrotating Killing field. In particular, dipole rings may be held in equilibrium even when their spin vanishes, thus demonstrating (infinite) non-uniqueness of magnetized static uncharged black holes in five dimensions. Physical properties of such rings are discussed.}

\keywords{Black Holes, Classical Theories of Gravity}

\preprint{gr-qc/0410048}

\begin{document}

\section{Introduction}

\label{sec_introduction}

In the past few years, there has been a significant increase in interest in the properties of gravity in more than four dimensions. This largely stems from the recognition of the relevance of black holes to fundamental theories such as string theory, along with the idea of large or infinite extra dimensions recently resurrected by TeV gravity models. Several higher dimensional solutions of classical General Relativity have been known for some time, in particular extensions to any $n>4$ of the Schwarzschild and Reissner-Nordstr\"{o}m black holes by Tangherlini \cite{Tangherlini63}, and of the Kerr black hole by Myers and Perry \cite{MyePer86}. However, recent investigations have shown that, even at the classical level, gravity in higher dimensions exhibits much richer dynamics than in $n=4$. One of the most intriguing features is the non-uniqueness of asymptotically flat rotating black holes. In five-dimensional vacuum General Relativity, explicit $S^1\times S^2$ rotating black ring solutions have been constructed \cite{EmpRea02prl} that may have the same mass and spin as the $S^3$ holes of \cite{MyePer86}. Such uniqueness violation in fact becomes continuously infinite for rings with magnetic ``dipole charge'' \cite{Emparan04}.

Analyses of uniqueness properties concern asymptotically flat
spacetime, a paradigm for isolated systems. However, external
fields tend to destroy asymptotic flatness. In $n=4$
Einstein-Maxwell theory, a ``uniform'' electromagnetic field is
described either by the Bertotti-Robinson family of direct product
geometries \cite{LeviCivita17BR,Bertotti59,Robinson59}, or by the
Melvin fluxtube \cite{Bonnor54,Melvin64}. Higher dimensional
magnetic extensions of the spacetimes
\cite{LeviCivita17BR,Bertotti59,Robinson59} are examples of
``spontaneous compactification'' \cite{FreRub80} (electric
counterparts emerge as extremal limits of static charged black
holes \cite{CarDiaLem04}). From an alternative point of view,
Melvin magnetic ``fluxbranes'' in $n\ge 4$ dimensions provide
brane world models with noncompact extra dimensions
\cite{Gibbons86,GibWil87}. Moreover, the embedding of such
fluxbranes in dilaton theories \cite{GibMae88} and the possibility
of obtaining them (in the Kaluza-Klein case) from a flat spacetime
with twisted identifications
\cite{Dowkeretal94_49,Dowkeretal94_50,Dowkeretal95,Dowkeretal96}
have opened the way for similar magnetic backgrounds in string
theory.

It is remarkable that in ${n=4}$ dimensions non-asymptotically
flat exact solutions of the Einstein-Maxwell equations exist that
describe black holes under the influence of external
electromagnetic fields. Ernst \cite{Ernst76a}  applied a Harrison
transformation \cite{Harrison68,Stephanibook} to the ``seed''
Schwarzschild  metric to elegantly obtain a static black hole in
the Melvin universe \cite{Bonnor54,Melvin64}. Various properties
of such Schwarzschild-Melvin solution have been subsequently
elucidated, e.g. in
\cite{WilKer80,Hiscock81,BosEst81,BicJan85,KarVok92,Radu02,Ortaggio04}.
More general magnetized Kerr-Newman metrics
\cite{Ernst76a,ErnWil76,GarciaD85} have provided exact models
where the ``coupling'' between rotation and magnetic fields gives
rise to interesting astrophysical effects, such as charge
accretion and flux expulsion from extreme holes
\cite{ErnWil76,Dokuchaev87,Karas88,AliGal89,KarVok91,KarBud00}
(discussed also in Kaluza-Klein and string theories
\cite{ChaEmpGib98}).

The purpose of the present paper is to study higher dimensional black holes in magnetic fields. We start by deriving the analogue of the Schwarzschild-Melvin solution
of \cite{Ernst76a} in any $n\ge 4$ spacetime dimension. In other
words, we will be considering a Schwarzschild-Tangherlini black
hole in an external magnetic field, represented by the Melvin
fluxbrane of \cite{Gibbons86,GibWil87}. Subsequently, we will 
comment on certain simple magnetized rotating solutions that do
not have any four-dimensional counterpart. For $n=5$, some of
these are related to recent results by Ida and Uchida
\cite{IdaUch03} and by Aliev and Frolov \cite{AliFro04} within
exact solutions and test fields approximation, respectively. We shall also analyze magnetized black holes with non-spherical topology, i.e. black rings (in $n=5$).  In particular, we will demonstrate that even static rings can be in equilibrium when they carry local dipole charge. 
We confine ourselves to the standard Einstein-Maxwell theory,
specified by the action~(\ref{action}) in Appendix~\ref{app_harrison}. The plan
of the paper is as follows. Following the method of
\cite{Ernst76a}, in Sec.~\ref{sec_magnetizing} we apply a
magnetizing Harrison transformation to the $n\ge 4$
Schwarzschild-Tangherlini line element. This results in a solution
of the Einstein-Maxwell equations representing a black hole
immersed in a ``uniform'' magnetic field, as discussed in
Sec.~\ref{sec_geometry}. We analyze how the Maxwell field deforms
the geometry of the event horizon by explicitly calculating the
associated Ricci scalar and the area of suitable spatial sections.
We notice that effects of flux concentration found in $n=4$
\cite{BicJan85} essentially occur in any dimension. Having in mind
recent studies of classical black hole production in high energy
scattering \cite{EarGid02,KohVen02,YosNam02}, we also perform the \AS boost of the magnetized
black hole. We thus obtain an impulsive gravitational wave
generated by a ``fast-moving'' particle in a magnetic field, which
generalizes previous results for ${n=4}$ \cite{Ortaggio04}. Both a
distributional and a continuous form of the corresponding line
element are presented. In Sec.~\ref{sec_rotating}, we observe that
for $n>4$ the simple magnetizing technique of
Sec.~\ref{sec_magnetizing} can be applied also to rotating
solutions, such as the Myers-Perry black hole and the
Emparan-Reall black ring. This is true provided there is at least
one nonrotating spacelike Killing vector, in which case one can
introduce a vector potential that is ``aligned'' (i.e., preserving
the symmetries of the original spacetime) but still nonrotating.
These simplified but exact models support conclusions from test
field approximations, according to which phenomena such as flux
expulsion arise only for ``rotating'' potentials. The relation of
our work with the previous studies \cite{IdaUch03,AliFro04} is
pointed out. Sec.~\ref{rings} analyzes in some detail the static limit of five-dimensional dipole rings held in equilibrium in a magnetic field. It is shown that there exists an infinite number of rings with the same mass and asymptotic magnetic field strength, which are labeled by the value of their local charge. Physical and thermodynamical quantities associated to these rings are computed. Appendix~\ref{app_harrison} reviews the $n\ge 4$ Harrison transformation employed in the paper and provides related references. An alternative expression for extremal static ring solutions which appeared originally in \cite{Emparan01npb} is given in Appendix~\ref{app_extreme}, together with the corresponding coordinate transformation. 

\section{Magnetizing the Schwarzschild-Tangherlini metric}

\label{sec_magnetizing}

A generalization of the Schwarzschild solution of the vacuum Einstein equations to spacetimes of arbitrary dimension $n\ge 4$ was found in \cite{Tangherlini63}. This is the spherically symmetric Schwarzschild-Tangherlini black hole,
which in hyperspherical coordinates takes the form
\begin{equation}
 \d s^2=-f^2\d t^2+f^{-2}\d r^2+r^2\d\Omega^2_{(n-2)} ,
 \label{tanghe1}
\end{equation}
where $\d\Omega^2_{(n-2)}$ is the standard line element on the unit $(n-2)$-sphere, and
\begin{equation}
 f^2=1-\frac{\mu}{r^{n-3}} .
 \label{gtt}
\end{equation}
The metric (\ref{tanghe1}) is asymptotically flat, and it has a
single spherical event horizon where $f=0$. The parameter $\mu>0$
is proportional to the physical mass $M$ \cite{MyePer86}
\begin{equation}
 M=\frac{\mu(n-2)\Omega_{n-2}}{16\pi} .
 \label{mass}
\end{equation}

Now we intend to study how the geometry (\ref{tanghe1}) is
modified when the black hole is not isolated but under the
influence of an external magnetic field. In the case of ${n=4}$
spacetime dimensions this was done by Ernst \cite{Ernst76a} by
means of a suitable Harrison transformation. It is shown in
Appendix~\ref{app_harrison} (see the original references therein) that the Harrison
transformation of \cite{Ernst76a}, based on the axial symmetry of
a seed solution, can be generalized to higher dimensions. Hence,
we can follow the same approach to obtain a higher dimensional
magnetized static black hole. Before doing that, it is convenient
to use the simple identity
$\d\Omega^2_{(n-2)}=\cos^2\theta\d\Omega^2_{(n-4)}+\d\theta^2+\sin^2\theta\d\phi^2$
in order to rewrite the Schwarzschild-Tangherlini
metric~(\ref{tanghe1}) as \beqn
 \d s^2=-f^2\d t^2+f^{-2}\d r^2+r^2\cos^2\theta\d\Omega^2_{(n-4)}+r^2\d\theta^2+r^2\sin^2\theta\d\phi^2 ,
 \label{tanghe2}
\eeqn 
in which $\theta\in[0,\pi/2]$, $\phi\in[0,2\pi]$ and
$\d\Omega^2_{(n-4)}=\d\psi_1^2+\sin^2\psi_1\d\psi_2+\ldots
+\prod_{a=1}^{n-5}\sin^2\psi_a\d\psi_{n-4}^2$ [except in the case
$n=4$, when $\d\Omega^2_{(n-4)}=0$, $\theta\in[0,\pi]$, and
Eqs.~(\ref{tanghe1}) and (\ref{tanghe2}) are of course
equivalent]. The line element (\ref{tanghe2}) is of the
form~(\ref{ansatzg}), and such that the squared norm of the
spacelike Killing vector $\pa_\phi$ takes the simple form
$V=g_{\phi\phi}=r^2\sin^2\theta$ independently of $n$. Since
Eq.~(\ref{tanghe2}) is a vacuum solution, we can use the
transformations (\ref{harrison}) with $A_\phi=0$ to generate a new
solution of the $n$-dimensional Einstein-Maxwell equations. The
transformed metric reads 
\be
 \d s^2=\Lambda^{2/(n-3)}\big[-f^2\d t^2+f^{-2}\d r^2+r^2\cos^2\theta\d\Omega^2_{(n-4)}+r^2\d\theta^2\big]+\Lambda^{-2}r^2\sin^2\theta\d\phi^2 ,
 \label{tanghemelvin}
\ee
with $f$ as in Eq.~(\ref{gtt}), and
\begin{equation}
 \Lambda=1+\frac{1}{2}\frac{n-3}{n-2}B^2r^2\sin^2\theta .
 \label{lambda_r}
\end{equation}
The associated vector potential and the corresponding magnetic field are given by (primes are dropped)
\beqn
 A & = & A_\phi\d\phi=\frac{1}{2}\Lambda^{-1}Br^2\sin^2\theta \d\phi , \label{potentialA} \\
 F & = & \Lambda^{-2}Br\sin\theta\left(\sin\theta\d r+r\cos\theta\d\theta\right)\wedge\d\phi \label{fieldF} .
\eeqn
For ${n=4}$ the results of \cite{Ernst76a} are recovered.\footnote{The $n=4$ Einstein-Maxwell theory is somewhat peculiar in that the electromagnetic 2-form field is only defined up to a constant duality rotation. The metric of \cite{Ernst76a} can thus be also associated to a purely electric 2-form (cf., e.g., \cite{Ortaggio04}). In general, the electric $(n-2)$-form dual of the magnetic field~(\ref{fieldF}) would provide a solution of the dual theory.} The constant $B$ introduced by the Harrison transformation parametrizes the strength of the magnetic field [the case $B=0$ simply corresponds to the original Schwarzschild-Tangherlini metric~(\ref{tanghe2}) with $A=0=F$]. In particular, the invariant
\begin{equation}
 \frac{1}{2}F^{\mu\nu}F_{\mu\nu}=\Lambda^{-2(n-2)/(n-3)}B^2(f^2\sin^2\theta+\cos^2\theta) ,
 \label{F2}
\end{equation}
takes the constant value $B^2$ at the ``axis'' $\theta=0$. The energy-momentum tensor $4\pi T_{\mu\nu}=F^\rho_{\ \, \mu}F_{\rho\nu}-\frac{1}{4}F^{\rho\sigma}F_{\rho\sigma}g_{\mu\nu}$ of the Maxwell field~(\ref{fieldF}) can be expressed using the orthonormal basis $\omega^0=\Lambda^{1/(n-3)}f\d t$, $\omega^1=\Lambda^{1/(n-3)}f^{-1}\d r$, $\omega^2=\Lambda^{1/(n-3)}r\d\theta$, $\omega^3=\Lambda^{-1}r\sin\theta\d\phi$, $\omega^4=\Lambda^{1/(n-3)}r\cos\theta\d\psi_1$,  $\omega^5=\Lambda^{1/(n-3)}r\cos\theta\sin\psi_1\d\psi_2$, $\ldots$, $\omega^{n-1}=\Lambda^{1/(n-3)}r\cos\theta\prod_{a=1}^{n-5}\sin\psi_a\d\psi_{n-4}$. The nonvanishing frame components are
\beqn
 8\pi T_{00} & = & \Lambda^{-2(n-2)/(n-3)}B^2(f^2\sin^2\theta+\cos^2\theta) , \nonumber \\
 8\pi T_{11} & = & \Lambda^{-2(n-2)/(n-3)}B^2(f^2\sin^2\theta-\cos^2\theta) , \nonumber \\
 4\pi T_{12} & = & \Lambda^{-2(n-2)/(n-3)}B^2f\sin\theta\cos\theta , \\
 T_{22} & = & -T_{11}, \quad T_{33}=-T_{44}=-T_{55}=\ldots=T_{00} . \nonumber
\eeqn

\section{Properties of the solution}

\label{sec_geometry}

\subsection{Black hole in Melvin background}

\label{subsec_bh}

The magnetized metric (\ref{tanghemelvin}) is static and
invariant, in particular, under rotations generated by $\pa_\phi$.
It has a single horizon located at $r=r_h\equiv\mu^{1/(n-3)}$
(where ${f=0}$), which is independent of the value of the magnetic
field strength $B$. As observed in \cite{Ernst76a,Hiscock81} for
the case $n=4$, the spacetime can be easily extended across the
horizon into a nonstatic region. In view of Eq.~(\ref{F2}), the
Ricci scalar
\begin{equation}
R=-\frac{16\pi}{n-2}T=\frac{n-4}{n-2}F^{\mu\nu}F_{\mu\nu} .
 \label{Riccisc}
\end{equation}
diverges at $r=0$, thus demonstrating the presence of a curvature
singularity.\footnote{Of course for $n=4$ one has $R=0=T$, but
there is still a curvature singularity inherited from the seed
Schwarzschild geometry, see, e.g., the Newman-Penrose scalars
calculated in \cite{BosEst81,Ortaggio04}.} On the other hand, for
$r\to\infty$ the line element (\ref{tanghemelvin}) approaches the
simpler form which one obtains by setting $f=1$ ($\mu=0$), i.e.
\be
 \d s_0^2=\Lambda^{2/(n-3)}\big[-\d t^2+\d r^2+r^2\cos^2\theta\d\Omega^2_{(n-4)}+r^2\d\theta^2\big]+\Lambda^{-2}r^2\sin^2\theta\d\phi^2 .
 \label{melvinradial}
\ee
Since ${\d r^2+r^2\d\theta^2=[\d(r\cos\theta)]^2+[\d(r\sin\theta)]^2}$, in this case it is convenient to replace the coordinates $\{r,\theta,\psi_1,\ldots,\psi_{n-4}\}$ by new coordinates $\{z_1,\ldots,z_{n-3},\rho\}$ satisfying
\be
 r\cos\theta=(z_1^2+z_2^2+\ldots+z_{n-3}^2)^{1/2} , \qquad 
 r\sin\theta=\rho ,
 \label{coord_z}
\ee
and such that  $[\d(r\cos\theta)]^2+(r\cos\theta)^2\d\Omega^2_{(n-4)}=\d z_1^2+\d z_2^2+\ldots+\d z_{n-3}^2$. Hence Eq.~(\ref{melvinradial}) can be rewritten as
\be
 \d s_0^2=\Lambda^{2/(n-3)}\big[-\d t^2+\d z_1^2+\d z_2^2+\ldots+\d z_{n-3}^2+\d\rho^2\big]+\Lambda^{-2}\rho^2\d\phi^2 ,
 \label{melvin}
\ee
whereas Eqs.~(\ref{lambda_r}), (\ref{potentialA}) and (\ref{fieldF}) become
\beqn
 \Lambda & = & 1+\frac{1}{2}\frac{n-3}{n-2}B^2\rho^2 , \label{lambda_rho} \\
 A & = & \frac{1}{2}\Lambda^{-1}B\rho^2\d\phi , \label{potentialAmelvin} \\
 F & = & \Lambda^{-2}B\rho\,\d\rho\wedge\d\phi \label{fieldFmelvin} .
\eeqn
The asymptotic solution given by Eqs.~(\ref{melvin})--(\ref{fieldFmelvin}) [equivalent to Eqs.~(\ref{tanghemelvin})--(\ref{fieldF}) with $\mu=0$] is the higher dimensional Melvin fluxbrane of \cite{Gibbons86,GibWil87} describing an ``originally uniform'' magnetic field which concentrates under its own gravity.\footnote{Note that the solution (\ref{melvin})--(\ref{fieldFmelvin}) of \cite{Gibbons86,GibWil87} can directly be obtained by applying the Harrison transformation of Appendix~\ref{app_harrison} to an $n$-dimensional Minkowski spacetime, given by Eq.~(\ref{melvin}) with $B=0$ (i.e., $\Lambda=1$; cf.~\cite{Ernst76a} for $n=4$). In the limit of a small $B$, from Eq.~(\ref{fieldFmelvin}) one obtains the solution $F=B\rho\,\d\rho\wedge\d\phi=B\d x\wedge\d y$ (where $x=\rho\cos\phi$, $y=\rho\sin\phi$) for a test uniform magnetic field on a Minkowski background.} With the previous observations, this suggest that we interpret the $n>4$ Einstein-Maxwell solution of Eqs.~(\ref{tanghemelvin})--(\ref{fieldF}) as a black hole in an external magnetic field, insomuch as the already investigated case of $n=4$ dimensions \cite{Ernst76a,Hiscock81}. The magnetized black hole (\ref{tanghemelvin}) is not asymptotically flat, but one can still compute its mass with the background subtraction method of \cite{HawHor96}. We easily find that the mass is unaffected by the magnetic field, and it is again given by Eq.~(\ref{mass}).

\subsection{Geometry of the horizon and thermodynamics}

\label{subsec_geometry}

It is interesting to analyze the effect of the magnetic field on the shape of the event horizon. The metric of  $(n-2)$-dimensional spatial sections of the horizon is
\begin{equation}
 \d  s^2_h=\Lambda_h^{2/(n-3)}r_h^2\big[\cos^2\theta\d\Omega^2_{(n-4)}+\d\theta^2\big]+
          \Lambda_h^{-2}r_h^2\sin^2\theta\d\phi^2 ,
 \label{horizon}
\end{equation}
where $\Lambda_h\equiv\Lambda|_{r=r_h}$. After straightforward calculations, the associated Ricci scalar
\beqn
 {\cal R}= & & \frac{1}{r_h^2}\frac{\Lambda_h^{-2(n-2)/(n-3)}}{n-2}\left\{(n-2)^2(n-3)\Lambda_h^2+
  2B^2r_h^2\left[n\cos^2\theta-(n-3)\sin^2\theta\right]\Lambda_h\right. \nonumber \\ 
  & & \left.{}-B^4r_h^4(n-1)\sin^2\theta\cos^2\theta\right\}  ,
   \label{Riccih}
\eeqn
provides us with a measure of the departure form sphericity in the
presence of a magnetic field. For ${B=0}$, Eq.~(\ref{Riccih})
reduces to ${{\cal R}=(n-2)(n-3)/r_h^2}$, since the horizon of the
Schwarzschild-Tangherlini spacetime (\ref{tanghe2}) is simply a
round $(n-2)$-sphere of radius $r_h$. For ${n=4}$ one recovers an
expression calculated in \cite{WilKer80}. Similarly as for the
discussion of \cite{EmpMye03} concerning the geometry of
(ultra\mbox{-)}spinning higher dimensional black holes, we can obtain
further invariant information by computing areas of privileged
sections of the horizon. Since the electromagnetic 2-form
(\ref{fieldF}) has only $F_{r\phi}$ and $F_{\theta\phi}$
components, it is natural to consider a ``parallel''
two-dimensional area obtained by fixing an arbitrary point on the
``transverse'' sphere $\Omega_{n-4}$. Integrating the square root
of the determinant
$\sqrt{\,g_{||}}=\Lambda_h^{-(n-4)/(n-3)}r_h^2\sin\theta$ one gets
[recall Eq.~(\ref{lambda_r})]
\begin{equation}
 {\cal A}_{||}^{(2)}=r_h^2\Omega_2\Lambda_0^{-(n-4)/(n-3)}
    F\left(\frac{n-4}{n-3},\frac{1}{2};\frac{3}{2};\frac{\Lambda_0-1}{\Lambda_0}\right) ,
\end{equation}
where $F$ is a hypergeometric function and $\Lambda_0\equiv\Lambda_h|_{\theta=\pi/2}$. For $B=0$ this expression simplifies to ${\cal A}_{||}^{(2)}=r_h^2\Omega_2$ (the same happens for $n=4$ \cite{WilKer80,BicJan85}, in which case ${\cal A}_{||}^{(2)}$ is the total area of the horizon). In general, ${\cal A}_{||}^{(2)}$ decreases with an increasing magnetic field $B$. From a complementary point of view, fixing $\theta,\phi=\mbox{constant}$ one can evaluate the area of a transverse sphere
\begin{equation}
 {\cal A}_{\bot}^{(n-4)}=\Lambda_h^{(n-4)/(n-3)}(r_h\cos\theta)^{n-4}\Omega_{n-4} .
\end{equation}
As opposed to ${\cal A}_{||}^{(2)}$, this area obviously
monotonically increases with $B$. Combining the above results, we
see that the horizon is ``pancaked'' along directions in the
transverse space $\Omega_{n-4}$, which  (conformally) expands
because of the magnetic field (note that the effect of a magnetic
field on the geometry of the horizon is thus ``opposite'' to that
due to rotation \cite{EmpMye03}; this was observed in
\cite{WilKer80} for $n=4$). However, deformations in the parallel
and transverse spaces conspire in such a way that the total area
of the event horizon is independent of the magnetic field. Namely,
\begin{equation}
 {\cal A}_h=r_h^{n-2}\Omega_{n-2} 
\end{equation}
is given by the same function of the mass~(\ref{mass}) (recall the comments at the very end of previous Subsec.~\ref{subsec_bh}) as in the case of the ``neutral'' Schwarzschild-Tangherlini metric~(\ref{tanghe2}). 

This physically interesting result was already known for $n=4$ \cite{WilKer80,BicJan85}. Mathematically, it is an obvious consequence of the Harrison transformation we have used to generate the metric (\ref{tanghemelvin}), which leaves the determinant $g_h$ of the line element~(\ref{horizon}) invariant. A similar invariance for the area of four-dimensional (composite) extreme black holes in magnetic fields was understood in \cite{Emparan97} in light of a microscopical interpretation of entropy. It is thus interesting to check whether other thermodynamical quantities are unaffected in the case of the magnetized Schwarzschild-Tangherlini black hole.  For $n=4$ this was done in \cite{Radu02}. The temperature can computed with standard techniques (Euclidean section or surface gravity) and one easily finds $\beta=1/T=4\pi r_h/(n-3)$. Since we already know that the mass (physical hamiltonian) is given by~(\ref{mass}), with the method of \cite{HawHor96} one obtains that the physical Euclidean action is $\tilde I_P=\beta M-\frac{1}{4}{\cal A}_h=\beta M/(n-2)$.\footnote{One can also calculate $\tilde I_P$ directly, using Eqs.~(\ref{F2}) and (\ref{Riccisc}), as done in four dimensions in \cite{Radu02}. For $n=4$, our results reduce to those of \cite{Radu02}.} Again, the dependence on the external magnetic field cancels out, and $\tilde I_P$ indeed coincides with the action computed in asymptotically flat spaces \cite{Dowkeretal96}. The standard area law for the entropy, $S=\frac{1}{4}{\cal A}_h$, follows readily. 

\subsection{Magnetic flux}

The amount of magnetic flux across a portion of the horizon provides a measure of how much the field~(\ref{fieldF}) threads the black hole. The flux through a closed curve $\gamma$ is given by the line integral $\Phi=\oint_\gamma A$. If we take $\gamma$ to be an orbit of the Killing field $\pa_\phi$ lying on the horizon, using Eq.~(\ref{potentialA}) we obtain for the corresponding flux
\begin{equation}
 \Phi=\frac{B\pi r_h^2\sin^2\theta}{1+\frac{1}{2}\frac{n-3}{n-2}B^2r_h^2\sin^2\theta} .
 \label{flux}
\end{equation}
This flux is maximum if the orbit $\gamma$ lies at $\theta=\pi/2$,
which for $n=4$ just corresponds to the boundary of the ``upper
half'' of the horizon \cite{BicJan85} [the factor $1/4$ in the
denominator of formula (41) of \cite{BicJan85} is incorrect,
cf.~\cite{Karas88,BicKar89}]. In Eq.~(\ref{flux}), the dependence on the
field strength $B$ is essentially the same for any $n$. As
observed in \cite{BicJan85} (see also \cite{Karas88,KarBud00}), by
increasing the parameter $B$ (with fixed $\mu$) the
flux~(\ref{flux}) first increases, as expected on classical
grounds. Then, it reaches its maximum value $\Phi_{\mbox{\small
max}}=\frac{\pi}{2}(2\frac{n-2}{n-3})^{1/2}r_h\sin\theta$ for
$B=B_{\mbox{\small
max}}=(2\frac{n-2}{n-3})^{1/2}(r_h\sin\theta)^{-1}$, and
eventually monotonically decreases, with $\Phi\to 0$ as
$B\to+\infty$. The existence of such an upper bound of the
magnetic flux is a relativistic effect caused by the concentration
of the field under its self-gravity. It disappears in the limit of
test fields, for which $\Phi_{\mbox{\small test}}=B\pi
r_h^2\sin^2\theta$ (cf., e.g., \cite{BicJan85,BicKar89}) is simply a linear function of $B$. 

\subsection{Ultrarelativistic limit}

\label{sec_boosting}

Extra-dimension models of TeV gravity have stimulated recent
investigations of classical black hole production in high energy
collisions in $n\ge4$ dimensions
\cite{EarGid02,KohVen02,YosNam02}. In such studies, the
gravitational field of each incoming particle is modelled as an
\AS impulse (or a modification of it), obtained by boosting a
Schwarzschild black hole to the speed of light in four
\cite{AicSex71} or higher \cite{LouSan90} dimensions. Recently we
applied an analogous ultrarelativistic boost to a $n=4$ black hole
immersed in a magnetic field, which resulted in an impulsive wave
propagating in the Melvin universe \cite{Ortaggio04}. In this
section we generalize the work \cite{Ortaggio04} to any $n\ge 4$.
In order to do that, we have to evaluate how the magnetized black
hole metric~(\ref{tanghemelvin}) transforms under an appropriate
Lorentz boost with velocity $V$, and perform the limit ${V\to 1}$.
Since for a large $r$ the line element~(\ref{tanghemelvin}) approaches the Melvin
spacetime (\ref{melvin}) [or (\ref{melvinradial})], a natural notion of boost is provided by the isometries of
the line element (\ref{melvin}), e.g. those generated by
$z_1\pa/\pa t+t\pa/\pa z_1$. The corresponding finite
transformation is simply expressed in terms of double null
coordinates
\begin{equation}
 u=\frac{z_1-t}{\sqrt{2}} , \qquad v=\frac{z_1+t}{\sqrt{2}} ,
 \label{nullcoords}
\end{equation}
as
\begin{equation}
 u\to A^{-1}u  , \qquad v\to Av  ,
 \label{lorentzboost}
\end{equation}
where $A>0$ is a parameter related to the standard Lorentz factor by $\gamma=(A+A^{-1})/2$. Before applying transformation (\ref{lorentzboost}) to the line element (\ref{tanghemelvin}), we decompose the latter as
\begin{equation}
 \d s^2=\d s_0^2+\Lambda^{2/(n-3)}\Delta ,
  \label{decomposition}
\end{equation}
in which $\d s^2_0$ is the Melvin spacetime (\ref{melvinradial}), and
\begin{equation}
 \Delta\equiv \mu\left(\frac{\d t^2}{r^{n-3}}+\frac{\d r^2}{r^{n-3}-\mu}\right) .
 \label{perturbation}
\end{equation}
Recalling the form (\ref{melvin}) of $\d s^2_0$, employing Eq.~(\ref{nullcoords}) and $r^2=(u+v)^2/2+z_2^2+\ldots+z_{n-3}^2+\rho^2$ [see Eq.~(\ref{coord_z})], Eqs.~(\ref{decomposition}) and (\ref{perturbation}) can be rewritten in coordinates $\{u,v,z_2,\ldots,z_{n-3},\rho,\phi\}$. One can thus make the substitution (\ref{lorentzboost}) in the transformed Eqs.~(\ref{decomposition}) and (\ref{perturbation}), which leaves $\d s^2_0$ and $\Lambda$ [see Eq.~(\ref{lambda_rho})] invariant and makes the quantity $\Delta\to\Delta_A$ dependent parametrically on $A$  (cf. \cite{Ortaggio04} for explicit expressions). After the standard rescaling \cite{AicSex71}
\begin{equation}
 M=2pA(1+A^2)^{-1} ,
 \label{ASmassrescaling}
\end{equation}
where $p>0$ is a constant, we study the ultrarelativistic limit $A\to 0$. The mathematics in the case $n>4$ is similar to that in $n=4$ \cite{Ortaggio04}, so we omit repetition of details here. We only observe that for $n>4$ no infinite ``gauge'' subtractions \cite{AicSex71,Ortaggio04} are required, and that the integral $2\int_0^\infty x^{2a-1}/(1+x^2)^{a+b}\d x=\Gamma(a)\Gamma(b)/\Gamma(a+b)$ has to be employed (with appropriate values $a,b>0$). After calculations, one finds that the ultrarelativistic limit $\d s^2=\d s_0^2+\Lambda^{2/(n-3)}\lim_{A\to 0}(\Delta_A)$ results in the final line element
\be
 \d s^2=\Lambda^{2/(n-3)}\big[2\d u\d v+\d z_2^2+\ldots+\d z_{n-3}^2+\d\rho^2\big]+\Lambda^{-2}\rho^2\d\phi^2+\Lambda^{2/(n-3)}H\delta(u)\d u^2 ,
 \label{melvinAS}
\ee
with
\beqn
 & H & = -8\sqrt{2}p\ln\rho \quad (n=4) , \label{H4} \\
 & H & = \frac{16\pi\sqrt{2}p}{(n-4)\Omega_{n-3}}\frac{1}{(z_2^2+\ldots+z_{n-3}^2+\rho^2)^{(n-4)/2}} \quad (n>4) .
 \label{Hn}
\eeqn
The spacetime (\ref{melvinAS}) simplifies to the form (\ref{melvin}) for ${u\neq 0}$. Accordingly, it represents an impulsive gravitational wave propagating in the Melvin background along the $z_1$ axis with the speed of light. The impulsive wave front, corresponding to the null hypersurface $u=0$, is not flat because of the background magnetic field. When the latter vanishes (for $B=0$, i.e., $\Lambda=1$), the metric (\ref{melvinAS}) reduces to the $n\ge 4$ \AS \pp wave \cite{AicSex71,LouSan90} in Minkowski spacetime. See \cite{Ortaggio04} for a more detailed analysis of the spacetime (\ref{melvinAS}) in the case $n=4$.

Thus far we have not considered the boost transformation of the Maxwell field (\ref{fieldF}) associated to the original, unboosted black hole (\ref{tanghemelvin}). Using Eq.~(\ref{coord_z}), the magnetic field (\ref{fieldF}) takes the form of Eq.~(\ref{fieldFmelvin}), which is clearly invariant under the boost (\ref{lorentzboost}). Therefore, the ultrarelativistic line element (\ref{melvinAS}) is a solution of the Einstein-Maxwell equations (except along the singular null line $u=0=z_2^2+\ldots+z_{n-3}^2+\rho^2$) with the magnetic field
\begin{equation}
 F=\Lambda^{-2}B\rho\,\d\rho\wedge\d\phi .
 \label{fieldAS}
\end{equation}
Indeed, one could alternatively obtain the solution~(\ref{melvinAS}), (\ref{fieldAS}) by directly applying the Harrison transformation~(\ref{harrison}) to the \AS vacuum spacetime \cite{AicSex71,LouSan90}.

In the studies \cite{EarGid02,KohVen02,YosNam02} of black hole formation in high energy collisions, it was convenient to employ an alternative form of the \AS metric that removes distributional terms. Therefore, we conclude this section presenting a new coordinate system $\{u,\tv,\tz_i,\tr,\phi\}$ (with $i=2,\ldots,n-3$) in which the metric (\ref{melvinAS}) contains only continuous functions. Namely, with the discontinuous substitution (sum $\sum_2^{n-3}$ is understood over the single index $l$ and over repeated indices $i,j,k$; $H_{,i}\equiv\pa H/\pa\tz_i$)
\beqn
 & & v=\tv-\frac{1}{2}\Theta(u)H-\frac{1}{8}u\Theta(u)\left(H_{,l}^2+H_{,\tr}^2\right) , \nonumber \\
 & & z_2=\tz_2+\frac{1}{2}u\Theta(u)H_{,2} ,  \nonumber \\
 & & \vdots  \label{disctransf} \\
 & & z_{n-3}=\tz_{n-3}+\frac{1}{2}u\Theta(u)H_{,n-3} ,  \nonumber \\
 & & \rho=\tr+\frac{1}{2}u\Theta(u)H_{,\tr} , \nonumber
\eeqn
one finds that Eq.~(\ref{melvinAS}) becomes
\beqn
 \d s^2= & & \Lambda^{2/(n-3)}\bigg\{2\d u\d\tv+\d\tz_2^2+\ldots+\d\tz_{n-3}^2+\d\tr^2+u\Theta(u)\left(H_{,ij}\d\tz_i\d\tz_j+2H_{,i\tr}\d\tz_i\d\tr+H_{,\tr\tr}\d\tr^2\right) \nonumber \\
        & & {}+\frac{1}{4}u^2\Theta(u)\big[\left(H_{,ik}H_{,kj}+H_{,i\tr}H_{,j\tr }\right)\d\tz_i\d\tz_j+\left(H_{,l\tr}^2+H_{,\tr\tr}^2\right)\d\tr^2 \nonumber \\  & & {}+2\left(H_{,ik}H_{,k\tr}+H_{,i\tr}H_{,\tr\tr}\right)\d\tz_i\d\tr\big]\bigg\}+\Lambda^{-2}\tr^2\left(1+\frac{1}{2}u\Theta(u)H_{,\tr}\tr^{-1}\right)^2\d\phi^2 .
\eeqn
Note that the transformation (\ref{disctransf}) is adapted to the present situation where $H$, given by Eqs.~(\ref{H4}) and (\ref{Hn}), is independent of $\phi$  (see \cite{Ortaggio04} for the case of a more general function $H$ in $n=4$).

\section{On rotating solutions}

\label{sec_rotating}

In the previous sections we studied an exact $n\ge 4$
Einstein-Maxwell solution describing a ``uniform'' magnetic field
threading a static black hole, obtained applying the Harrison
transformation of Appendix~\ref{app_harrison}. A natural next step would be to
extend our investigation to rotating black holes. In $n=4$, the
construction of Kerr-Newman black holes in magnetic fields
\cite{Ernst76a,ErnWil76,GarciaD85} required a Harrison
transformation more general and complex than the one
considered in the present paper [because the seed Kerr-Newman
metric is not of the form~(\ref{ansatzg})]. A systematic study of
rotating magnetized black holes in higher dimensions goes beyond
the scope of this work, and it is left for future investigations.
Nevertheless, it is worth remarking here that the simple Harrison
transformation employed in Sec.~\ref{sec_magnetizing} may be used
to generate some (special) magnetized solutions also in the
presence of rotation, provided $n>4$.

\subsection{Magnetized black holes}

The Myers-Perry line element \cite{MyePer86} is the natural generalization of the Kerr solution in $n>4$ dimensions. It admits $\lfloor (n-1)/2\rfloor\ge 2$ commuting spatial Killing vectors associated with independent rotations in orthogonal planes (the symbol $\lfloor \, \rfloor$ denotes integer part). If one (but not all) of the spin parameters is set to zero, the $n>4$ metric of \cite{MyePer86} is still rotating but does take the form~(\ref{ansatzg}). Accordingly, it can be immersed in an external magnetic field with the method described in Appendix~\ref{app_harrison}. For the sake of definiteness, let us present explicitly the corresponding magnetized Myers-Perry spacetime in the case of odd $n$ (the case of even $n$ works similarly). Applying the Harrison transformation~(\ref{harrison}) to the  solution of \cite{MyePer86} with a vanishing spin $a_1=0$ (say), one obtains the metric
\beqn
 \d s^2= & & \Lambda^{2/(n-3)}\bigg[-\d t^2+(r^2+a_i^2)(\d\mu_i^2+\mu_i^2\d\phi_i^2)+\frac{\mu r^2}{\Pi {\cal F}}(\d t+a_i\mu_i^2\d\phi_i)^2 \nonumber \label{MP} \\ & & {}+\frac{\Pi {\cal F}}{\Pi-\mu r^2}\d r^2+r^2\d\mu_1^2\bigg]+\Lambda^{-2}r^2\mu_1^2\d\phi_1^2 ,
\eeqn
where sum over ${i=2,\dots,(n-1)/2}$ is understood. The direction cosines satisfy $\mu_1^2+\sum_{i=2}^{(n-1)/2}\mu_i^2=1$, $\mu$ and $a_2,\ldots,a_{(n-1)/2}$ are constants related to the mass and angular momenta, $\Pi(r)$ and ${\cal F}(r,\mu_i)$ are the standard functions of \cite{MyePer86}, and
\begin{equation}
 \Lambda=1+\frac{1}{2}\frac{n-3}{n-2}B^2r^2\mu_1^2 .
\end{equation}
The vector potential and Maxwell field are
\beqn
 A & = &  A_{\phi_1}\d\phi_1=\frac{1}{2}\Lambda^{-1}Br^2\mu_1^2\d\phi_1 , \label{potentialMP} \\
 F & = & \Lambda^{-2}Br\mu_1\left(\mu_1\d r+r\d\mu_1\right)\wedge\d\phi_1 \label{fieldMP}.
\eeqn
For $n=5$, one recovers a solution presented in \cite{IdaUch03}. To the linear order in $B$, the latter describes test magnetic fields on the $n=5$ Myers-Perry background, studied in detail very recently \cite{AliFro04} (without the special requirement $a_1=0$, and with a vector potential represented by an arbitrary combination of all the three Killing vectors). Note that in general, by construction, the vector potential~(\ref{potentialMP}) points along the only nonrotating Killing vector. This is of course a simplifying assumption, not possible in $n=4$. As a consequence, the 2-form field~(\ref{fieldMP}) is purely magnetic (at least for locally nonrotating observers) and has no associated electric charge. Moreover, the potential~(\ref{potentialMP}) is independent of the rotation parameters $a_i$. These observations should be contrasted with the complex physical effects displayed by the $n=4$ solutions of \cite{ErnWil76,GarciaD85}, as analyzed in \cite{Dokuchaev87,Karas88,AliGal89,KarVok91,KarBud00}. Nevertheless, they support results from test field approximations according to which phenomena such as flux expulsion are connected to vector potentials having components in rotating planes \cite{BicJan85,AliFro04}.

\subsection{Magnetized black (dipole) rings}

In $n=5$ dimensions, the Myers-Perry solution does not represent the unique asymptotically flat rotating black hole. There exist also rings with a $S^1\times S^2$ horizon \cite{EmpRea02prl}, possibly carrying ``local'' magnetic charge (and with an arbitrary dilaton coupling, which we will set to zero) \cite{Emparan04}. By construction these rings are of the form~(\ref{ansatzg}), as they rotate in a single plane. Therefore we can again add an external magnetic field employing the transformation~(\ref{harrison}).\footnote{Since the seed solution of \cite{Emparan04} itself contains a nonvanishing seed electromagnetic potential $A=A_\phi\d\phi$, in this subsection (and only here) we adopt the full notation of~(\ref{harrison}) with a primate index for the transformed vector potential $A'=A'_\phi\d\phi$.} The metric of \cite{Emparan04} thus becomes
\beqn
 \d s^2= & & -\Lambda\frac{{\cal F}(y)}{{\cal F}(x)}\frac{{\cal H}(x)}{{\cal H}(y)}\left(\d t+C_1L\frac{1+y}{{\cal F}(y)}\d\psi\right)^2+\Lambda L^2\frac{{\cal F}(x){\cal H}(x){\cal H}(y)^2}{(x-y)^2} \nonumber \label{ring} \\
 & & {}\times\left[\frac{-{\cal G}(y)}{{\cal F}(y){\cal
     H}(y)^3}\d\psi^2-\frac{\d y^2}{{\cal G}(y)}+\frac{\d x^2}{{\cal G}(x)}+\Lambda^{-3}\frac{{\cal G}(x)}{{\cal
   F}(x){\cal H}(x)^3}\d\phi^2\right] ,
\eeqn
where
\beqn
 \Lambda=\left(1+\frac{2}{3}BA_\phi\right)^2+\frac{1}{3}B^2L^2\frac{{\cal H}(y)^2{\cal G}(x)}{(x-y)^2{\cal H}(x)^2} .
 \label{Lambdaring}
\eeqn
The remaining functions come from the seed metric and the associated seed vector potential~\cite{Emparan04}
\beqn
 {\cal F}(\xi) & = & 1+\lambda\xi , \qquad {\cal G}(\xi)=(1-\xi^2)(1+\nu\xi), \nonumber \\
 {\cal H}(\xi) & = & 1-\mu\xi , \qquad A_\phi=\frac{1}{2}C_2\sqrt{3}L\frac{1+x}{{\cal H}(x)} ,
\eeqn
and we take $x\in[-1,1]$ and $y\in(-\infty,-1]\cup (1/\mu,+\infty)$ ($y=-1/\nu$ and $y\to\infty$ are horizons, $y=-1/\lambda$ an ergosurface, and $y=1/\mu$ a curvature singularity \cite{Emparan04}). The constant $L>0$ is related to the radius of the ring, whereas $C_1,C_2>0$ are expressible in terms of the dimensionless parameters $\lambda,\nu,\mu$ (see \cite{Emparan04} for details)
\beqn
 & & C_1=\sqrt{\lambda(\lambda-\nu)\frac{1+\lambda}{1-\lambda}} , \qquad C_2=\sqrt{\mu(\mu+\nu)\frac{1-\mu}{1+\mu}} \nonumber \label{parameters} \\
  & & 0<\nu\le\lambda<1 , \qquad 0\le\mu<1 .
\eeqn
The potential associated to the new metric~(\ref{ring}) is
\begin{equation}
 A'=\Lambda^{-1}\left[A_\phi+B\left(\frac{1}{2}L^2\frac{{\cal H}(y)^2{\cal G}(x)}{(x-y)^2{\cal
 H}(x)^2}+\frac{2}{3}A_\phi^2\right)\right]\d\phi .
 \label{potential_ring}
\end{equation}
To avoid conical singularities at the axes $x=-1$ and $y=-1$ the angular coordinates must have periodicity \cite{Emparan04}
\be
 \Delta\phi=2\pi\frac{(1+\mu)^{3/2}\sqrt{1-\lambda}}{1-\nu}=\Delta\psi . 
 \label{periodicity}
\end{equation}
Forces acting on the black ring are in balance if conical singularities are absent also at $x=+1$, which constraints the five parameters of the solution~(\ref{ring})
\begin{equation}
 \frac{1+\lambda}{1-\lambda}\left(\frac{1-\mu}{1+\mu}\right)^3\Lambda^3|_{x=+1}=\left(\frac{1+\nu}{1-\nu}\right)^2 .
 \label{equilibrium}
\end{equation}
The presence of the parameter $B$ (via $\Lambda$) in the above equilibrium condition manifests the coupling of the magnetic charge to the external magnetic field (for $B=0$ one recovers the condition of \cite{Emparan04}). Notice, however, that $B$ does not represent here the physical field strength defined asymptotically (see, e.g., \cite{Dowkeretal94_49,Emparan97}). Indeed, for $x\to y\to-1$ the line element~(\ref{ring}) asymptotes the $n=5$ Melvin fluxbrane~(\ref{melvin}) (after a suitable coordinate transformation/rescaling, cf.~\cite{Dowkeretal94_50,EmpRea02prd}) with field strength
\begin{equation}
 B_0=\frac{1-\nu}{(1+\mu)^{3/2}\sqrt{1-\lambda}}B .
 \label{Bring}
\end{equation}
The local charge \cite{Dowkeretal96,Emparan04} of the ring is
\begin{equation}
 {\cal Q}=\frac{1}{2}\sqrt{3}L\frac{1+\mu}{1-\nu}\sqrt{\frac{\mu(\mu+\nu)(1-\lambda)}{1-\mu}}\Lambda^{-1/2}|_{x=+1} .
 \label{charge}
\end{equation}

The solution (\ref{ring}), (\ref{potential_ring}) admits various interesting limits for specific choices of the parameters. For $\mu=0$ (i.e., ${\cal Q}=0$) one has a magnetized version of the neutral rotating ring of \cite{EmpRea02prl}, in which centrifugal repulsion balances gravitational self-attraction. On the other hand, if $\lambda=\nu$ the spacetime~(\ref{ring}) becomes static, yet equilibrium is possible (if $\mu\neq 0$) thanks to the interaction between the local charge and the external magnetic field (see the next section). If we set simultaneously $\mu=0$ and $\lambda=\nu$, we obtain the neutral static ring of \cite{EmpRea02prd} immersed in a magnetic field, which can not be in equilibrium due to its unbalanced self-gravity. For $\lambda=\nu=\mu=0$, Eqs.~(\ref{ring}), (\ref{potential_ring}) simply describe a five-dimensional Melvin fluxbrane in unusual coordinates. The spacetime is of course flat if, in addition, $B=0$ (it can be put in standard form with a transformation given in \cite{EmpRea02prd}).

\section{Static rings in equilibrium}

\label{rings}

As mentioned above, the dipole rings of \cite{Emparan04} can be held in equilibrium also in the {\em static limit} $\nu=\lambda$, provided one switches on a magnetic field with an appropriate strength. In this section we analyze various physical properties of such special configurations, for which $C_1=0$ and ${\cal G}(\xi)=(1-\xi^2)F(\xi)$.The metric of magnetized static rings (\ref{ring}) thus simplifies to 
\beqn
 \d s^2= & & -\Lambda\frac{{\cal F}(y)}{{\cal F}(x)}\frac{{\cal H}(x)}{{\cal H}(y)}\d t^2+\Lambda L^2\frac{{\cal F}(x){\cal H}(x){\cal H}(y)^2}{(x-y)^2} \nonumber \label{staticring} \\
 & & {}\times\left[\frac{y^2-1}{{\cal H}(y)^3}\d\psi^2+\frac{\d y^2}{(y^2-1){\cal F}(y)}+\frac{\d x^2}{(1-x^2){\cal F}(x)}+\Lambda^{-3}\frac{1-x^2}{{\cal H}(x)^3}\d\phi^2\right] ,
\eeqn
while the factor (\ref{Lambdaring}) and the vector potential (\ref{potential_ring}) are essentially unchanged. Balance between gravitational and electromagnetic forces is achieved if [cf. Eqs.~(\ref{periodicity}) and (\ref{equilibrium})]
\be
\Delta\phi=2\pi\frac{(1+\mu)^{3/2}}{\sqrt{1-\lambda}}=\Delta\psi , \qquad     
  \left(\frac{1-\mu}{1+\mu}\right)^3\Lambda^3|_{x=+1}=\frac{1+\lambda}{1-\lambda} . 
 \label{equilibrium_static}
\ee
It is easy to see that the second of these equation can always be solved to determine $B$ as a function of arbitrarily specified $\lambda$ and $\mu$ [in the range allowed by Eq.~(\ref{parameters})]. These specific values of $B$ exactly cancels the conical singularity (in the form of a deficit/excess membrane) that is necessary to support static rings when $B=0$ \cite{EmpRea02prd,Emparan04}. One can thus have five-dimensional static black holes with a regular horizon of non-spherical topology (and therefore different from the solution studied in Secs.~\ref{sec_magnetizing} and \ref{sec_geometry}). This was first realized in \cite{Emparan01npb} for the case of extremal black holes with a regular but degenerate horizon (corresponding to $\lambda=0$, see also Appendix~\ref{app_extreme}). In fact, away from extremality there exist a continuous infinity of rings with the same mass and asymptotic magnetic field that are distinguished by the parameter~${\cal Q}$ (this will be detailed below), which is not an asymptotically conserved charge \cite{Emparan04}. Such non-uniqueness of ${n=5}$ asymptotically Melvin, static, (globally) uncharged black holes should be contrasted with the uniqueness of the asymptotically flat Schwarzschild-Tangherlini solution in $n\ge 4$ \cite{Hwang98,GibIdaShi02prl,GibIdaShi02prd} and of the Schwarzschild-Melvin solution in $n=4$ \cite{Hiscock81}.\footnote{I am thankful to Roberto Emparan for suggesting that I should emphasize this point, and for related useful remarks.} 

So far it has been technically convenient to specify the black ring spacetime in terms of the dimensionless parameters $\lambda$ and $\mu$, the ``radius'' $L$ and the Harrison-transformation constant $B$ [not all independent because of the second of Eqs.~(\ref{equilibrium_static})]. Now we will rather characterize balanced dipole rings in terms of the physical quantities $(M,B_0,{\cal Q})$, i.e., their mass, asymptotic field strength [see Eq.~(\ref{Bring})] and local charge. We will also briefly comments on their thermodynamics.

\subsection{Mass}

Although the static black ring is not asymptotically flat, its total energy can be defined with respect to a suitable static background \cite{HawHor96}. Namely, the black ring mass is given by the value of the physical hamiltonian
\be
 M=H_P=-\frac{1}{8\pi}\int N(^3K-\,{^3K}_0) ,
 \label{energy} 
\ee
where the integral is over a (three-dimensional) spatial boundary ``near infinity'', $N$ is the lapse, $^3K$ is the trace of the extrinsic curvature of the boundary as embedded in a spacelike slice of constant~$t$, and $^3K_0$ is the analogous quantity for the background spacetime. In the case of the black ring (\ref{staticring}), the reference background is the five-dimensional Melvin fluxbrane [obtained by setting $\lambda=\mu=0$ in Eq.~(\ref{staticring})], whereas $\pa_t$ is the Killing vector appropriately normalized on the axis $x=-1$ at infinity. In order to calculate the integral~(\ref{energy}), we need to take a boundary near infinity, calculate its extrinsic curvature, and eventually consider the limit as the boundary goes to infinity. Since in Eq.~(\ref{energy}) there is one term for the black ring and one for the background, we have to make sure that the intrinsic geometry and the Maxwell field on the two boundaries that we use are the same (to a ``sufficient'' order \cite{HawHor96}). 
Following a procedure used in a similar calculation in \cite{HawHorRos95}, near infinity (i.e., $x\to y\to-1$) we assume a boundary of the form 
\beqn
 x & = & -1+\e(1+\mu)^3\chi[1+\e(k_1\chi+k_2)] , \nonumber \\
 y & = & -1+\e(1+\mu)^3(\chi-1)[1+\e(k_1\chi+k_3)] , 
 \label{boundary}
\eeqn
where
\beqn
 k_1 & = & \frac{\lambda}{1-\lambda}+\frac{\mu\left[3(1+\lambda)+3(5-3\lambda)\mu+(9-7\lambda)\mu^2\right]}{2(1-\lambda)} , \nonumber \\
 k_2 & = & -\mu(1+5\mu+3\mu^2) , \label{k123} \\
 k_3 & = & \frac{1}{2}(1+\mu^2)(1-\mu) . \nonumber
\eeqn
The limit $\e\to 0$ corresponds to infinity, $\chi\in[0,1]$ being a convenient coordinate there. Using Eq.~(\ref{boundary}) and defining new angles $\psi_0,\phi_0\in[0,2\pi]$ 
\be
 \psi_0=\frac{\sqrt{1-\lambda}}{(1+\mu)^{3/2}}\psi , \quad \phi_0=\frac{\sqrt{1-\lambda}}{(1+\mu)^{3/2}}\phi ,
\ee 
the intrinsic metric induced on the boundary is
\be
 \d s^2=\Lambda\frac{2L^2}{\e}\bigg[\left(1+\e\frac{3\chi-2}{2}\right)\frac{\d\chi^2}{4\chi(1-\chi)}  
 +(1-\chi)(1+\e\chi)\d\psi_0^2+\Lambda^{-3}\chi[1+\e(\chi-2)]\d\phi_0^2\bigg] ,
 \label{boundaryg}
\ee
where
\be
 \Lambda=\frac{2L^2}{3\e}\frac{1-\lambda}{(1+\mu)^3}B^2\chi+1+\frac{L^2}{3}\frac{1-\lambda}{(1+\mu)^3}B^2\chi (\chi-2) ,
 \label{boundaryLambda}
\ee
and all quantities are evaluated to second nontrivial order in $\e$ (higher order terms will not contribute in the limit $\e\to 0$). The magnetic field associated to the potential~(\ref{potential_ring}) on the boundary is
\be
 F=\frac{9\e}{4L^2}\frac{(1+\mu)^{9/2}}{(1-\lambda)^{3/2}}\frac{1}{B^3}\frac{\d\chi\wedge\d\phi_0}{\chi^2}
 \left[1+\e\left(1-\frac{3}{L^2\chi}\frac{(1+\mu)^{3}}{1-\lambda}\frac{1}{B^2}\right)\right] .
 \label{boundaryF}
\ee
It is now evident that the boundary fields~(\ref{boundaryg}) [with Eq.~(\ref{boundaryLambda})] and (\ref{boundaryF}) do match with the corresponding quantities calculated for a five-dimensional Melvin fluxtube, provided the latter has field strength~(\ref{Bring}) (recall $\nu=\lambda$ here).\footnote{In fact, it is just by requiring such a matching that we found the specific values of the parameters $k_1,k_2,k_3$ in Eq.~(\ref{k123}).}

Once the boundaries are matched, we can proceed calculating the extrinsic curvature of the boundary~(\ref{boundaryg}), and similarly for the background. Taking the difference, divergent terms cancel out and one is left with ($^3h$ is the determinant of the 3-metric) 
\beqn
 \sqrt{^3h}N(^3K-\,{^3K}_0)=-\frac{3L^2}{2}(1+\mu)^2\frac{\lambda+\mu}{1-\lambda} .
\eeqn
Plugging this into the definition~(\ref{energy}), we obtain the ring mass
\be
 M=\frac{3\pi L^2}{4}\frac{(1+\mu)^2(\lambda+\mu)}{1-\lambda} , 
 \label{ringmass}
\ee
which does not depend explicitly on the background magnetic field, and indeed it coincides with that of the asymptotically flat solution of~\cite{Emparan04} [but note that the condition~(\ref{equilibrium}) does involve $B$].

\subsection{Local charge and horizon area}

Using Eqs.~(\ref{Lambdaring}), (\ref{Bring}) and (\ref{ringmass}), we can rewrite the local charge~(\ref{charge}) as
\be
 {\cal Q}=\left(\sqrt{\frac{\pi}{M}}\sqrt{\frac{1-\mu}{\mu}}+\frac{4}{3}B_0\right)^{-1} , 
 \label{charge2}
\ee
which is a growing function of $\mu$ restricted to ${\cal Q}\in\left[0,3/(4B_0)\right)$. One can easily invert this relation to find $\mu$ as a function of $(M,B_0,{\cal Q})$. 

With Eq.~(\ref{ringmass}), the area of the outer horizon $y=-1/\lambda$ reads 
\be
 {\cal A}_h=\frac{64}{3}\sqrt{\frac{\pi}{3}}\sqrt\frac{\lambda}{1+\lambda}M^{3/2} .
 \label{ringarea}
\ee
One can use the constraint~(\ref{equilibrium_static}) to get rid of $\lambda$, and use the inverse of Eq.~(\ref{charge2}) to eventually express ${\cal A}_h$ as a function of the physical parameters $(M,B_0,{\cal Q})$ only. Evidently, there exist an infinite number of static black rings with the same mass $M$ and asymptotical magnetic field $B_0$, which are labeled by ${\cal Q}$ (recall that ${\cal Q }$ is {\em not} a conserved asymptotic charge \cite{Emparan04}).  This resembles the non-uniqueness of asymptotically flat rotating dipole rings with given mass and angular momentum \cite{Emparan04}. Here, we can explore the multiplicity of magnetized static solutions by studying how their horizon area varies with ${\cal Q}$, keeping $M$ and $B_0$ fixed. For this purpose, it is convenient to follow \cite{Emparan04} in introducing dimensionless magnitudes 
\be
 b_0\equiv \sqrt{M}B_0, \qquad q\equiv\frac{{\cal Q}}{\sqrt{M}} , \qquad a_h\equiv\frac{3}{64}\sqrt{\frac{6}{\pi}}\frac{{\cal A}_h}{M^{3/2}} ,
 \label{dimensionless} 
\ee
so that $q\in\left[0,3/(4b_0)\right)$. For rings of given mass, the reduced area can be written in terms of $(b_0,q)$ as
\be
 a_h=\sqrt{1-\frac{1}{729}\left[18\pi q^2+(3-4b_0q)^2\right]^3} ,
 \label{area}
\ee
and it is plotted in Fig.~\ref{fig_area} as a function of $q$, for different values of $b_0$.
\FIGURE[ht]{
\epsfig{file=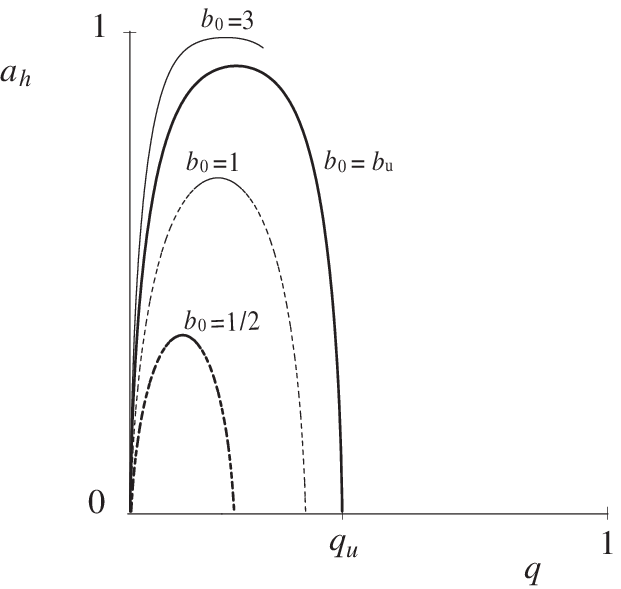,width=8.7cm}
\caption{Plot of the horizon area~(\ref{area}) as a function of the local charge $q$, for static black rings with a given mass. The four curves refer to different representative values of the asymptotic field strength $b_0>0$. The possible range of $q$ depends on $b_0$: $q\in[0,q_*]$ for $b_0<b_u$, and  $q\in\left[0,3/(4b_0)\right)$ otherwise. The values $q=q_*>0$ ($q_*$ depends on $b_0$) correspond to intersections with the $q$-axes and represent extremal rings with zero horizon area. For $b_0=b_u$, $q$ can take values arbitrarily close to the absolute upper bound $q_u\equiv(2\pi)^{-1/2}$, without reaching it. For any given $b_0$, $a_h$ takes its maximum value when $q=q_*/2$. We have chosen the normalization~(\ref{dimensionless}) for $a_h$ so that such a maximum approaches $1$ for $b_0\to\infty$. All the curves intersect at the origin of the axes, which simply describes the Melvin background.}
\label{fig_area}}
Notice that $a_h=0$ either for $q=0$, which is simply the Melvin background (equilibrium has been already enforced)\footnote{As long as we insist that $M$ is constant and that forces on the ring are in balance, the limit ${\cal Q}\to 0$ leads to a singular metric in the coordinate system used so far ($L$ blows up as ${\cal Q}^{-1/2}$). Therefore, one should perform a transformation very similar to the one used in the black string limit in \cite{Emparan04} (which in turn resembles the limit of zero acceleration in the well known $C$-metric).}, or for $q=q_*(b_0)\equiv12b_0(9\pi+8b_0^2)^{-1}$, which corresponds to the extremal rings of \cite{Emparan01npb}. 
Such extremal configurations are possible, however, only when $b_0<b_u\equiv(9\pi/8)^{1/2}$ [so that $q_*<3/(4b_0)$], in which case $q$ has to be further restricted to $q\in[0,q_*]$ (this follows from the equilibrium condition, and it also ensures that $a_h$ is real). 

\subsection{Temperature, Euclidean action and entropy}

We finally discuss thermodynamical properties of the static ring. The temperature can be straightforwardly determined by taking the Euclidean section and requiring regularity of the Euclidean continuation of the solution (\ref{staticring}) at the horizon (or, equivalently, from the surface gravity definition). One finds
\be
 T=\frac{1}{\beta}=\frac{1}{4\pi L}\frac{\sqrt{\lambda(1-\lambda^2)}}{(\lambda+\mu)^{3/2}} , 
\ee
which does not contain the parameter $B$ and agrees with the result obtained in \cite{Emparan04} for the case $B=0$. 

In order to compute the Euclidean action, we follow again the background subtraction method of \cite{HawHor96} (cf. also \cite{HawHorRos95}), and we define the physical action with respect to the Melvin background. Since we have already kept into account the background contribution in the calculation of the physical hamiltonian $H_P(=M)$ performed above, we can write the physical Euclidean action directly as $\tilde I_P=\beta H_P-\frac{1}{8\pi}\int K$,
where the integral is over a (four-dimensional) small neighbourhood of the outer horizon, and $K$ is the trace of the extrinsic curvature of such a boundary. Taking the outward unit normal $n_\mu=\Lambda^{1/2}L{\cal F}(x)^{1/2}{\cal H}(x)^{1/2}{\cal H}(y)/((x-y)\sqrt{y^2-1}{\cal F}(y)^{1/2})\delta_\mu^y$, the induced metric is $h_{\mu\nu}=g_{\mu\nu}-n_\mu n_\nu$, and $K=h^{\mu\nu}n_{\mu;\nu}$. Using the specific form of Euclidean solution corresponding to Eq.~(\ref{staticring}), we can perform the integration explicitly and obtain
\be
 \tilde I_P=\beta H_P-\frac{1}{4}{\cal A}_h . 
 \label{EuclideanI}
\ee
Terms depending on $B$ cancel out during the calculation [recall that the horizon area (\ref{ringarea}) does not contain $B$]. In the standard semiclassical approximation $\log Z\approx -\tilde I_P$ \cite{HawHor96,HawHorRos95}, from Eq.~(\ref{EuclideanI}) one finds that the entropy $S=-(\beta\pa_\beta-1)\log Z$ satisfies the area law $S=\frac{1}{4}{\cal A}_h$. For asymptotically flat neutral rings ($B_0=0={\cal Q}$) this result was found in \cite{EmpRea02prd}.

\section{Conclusions}

We have generalized to any $n\ge 4$ the Ernst \cite{Ernst76a} construction of static black holes in a magnetic field, which relies on using a Harrison transformation. We have discussed physical and geometrical properties of the solution, such as the geometry of the event horizon, the behaviour of magnetic flux, and the \AS ultrarelativistic limit of the spacetime. Most of these results are extensions of previously known facts in four dimensions. However, we have also considered rotating solutions (such as the Myers-Perry black hole and the Emparan-Reall black ring) when one of the spins vanishes. In this case, one can generate magnetized solutions that do not have any four-dimensional counterpart. Although simplified, these models confirm the expectation (based on intuition and on results for test fields in $n=5$ \cite{AliFro04}) that rotation and magnetic field ``do not couple'' if the vector potential is parallel to a  nonrotating Killing field. Moreover, we have shown that magnetized dipole rings may be held in equilibrium even in the limit of zero rotation. They thus provide infinite examples of static, regular black holes different from the magnetized Schwarzschild-Tangherlini spacetime, but which can have the same mass and asymptotics. 

Further study should possibly focus on a more general higher dimensional Harrison transformation employing a rotating Killing vector, that is an ansatz more general than Eq.~(\ref{ansatzg}). Similarly, one could consider a Harrison transformation in which the seed and the transformed vector potential are no longer aligned [as it is in Eq.~(\ref{harrison})]. These extensions, following up on \cite{Ernst76a,ErnWil76,GarciaD85}, would enable one to magnetize rotating solutions with rotating vector potentials, as well as the Reissner-Nordstr\"{o}m black holes of \cite{Tangherlini63}, for example. Eventually, one could generalize to $n>4$ the study of the rich phenomenology of $n=4$ dimensions
\cite{ErnWil76,Dokuchaev87,Karas88,AliGal89,KarVok91,KarBud00} and, within exact models, find a $n\ge 5$ counterpart of the interesting test field results of \cite{AliFro04} in $n=5$. It is worth remarking that Harrison transformations exist also for ``effective'' theories with scalar and additional gauge fields \cite{Dowkeretal94_49,Dowkeretal94_50,Dowkeretal95,Dowkeretal96,Ross94,Emparan97,GalRyt98,Emparan01npb}, which are relevant to superstring and supergravity theories. One could therefore extend the analysis of the present paper beyond Einstein-Maxwell theory. See, e.g., \cite{GibMae88,Dowkeretal96,ChaEmpGib98,GalRyt98,Emparan01npb} for related results in higher dimensions.

Finally, it would be interesting to analyze the ultrarelativistic limit of the black rings considered above. Although this is in principle analogous to the \AS boost of the magnetized Schwarzschild-Tangherlini black hole studied in this paper, it turns out to be technically more complex. A detailed study of a lightlike boost in the case of the static neutral rings of \cite{EmpRea02prd} has been recently presented in \cite{OrtKrtPod05}.

\acknowledgments

I wish to thank Roberto Emparan for inspiring suggestions at an early stage of this work, and for many helpful  comments on the draft. I am also grateful to 
Ji\v{r}\'{\i} Bi\v c\'ak and Ji\v{r}\'{\i} Podolsk\'y for reading the manuscript. The author is supported by a post-doctoral fellowship from Istituto Nazionale di Fisica Nucleare (bando n.10068/03).

\appendix

\section{Harrison transformation}
\label{app_harrison}

Harrison \cite{Harrison68} (see also \cite{Stephanibook} and
references therein) investigated systematic methods to generate
new solutions of the Einstein-Maxwell equations from old ones in
$n=4$ spacetime dimension, relying on the presence of a nonnull
Killing vector field. A Harrison-type transformation was presented
in \cite{Dowkeretal94_49} that generates background magnetic
fields in Einstein-Maxwell-scalar theories with an arbitrary
dilaton coupling. A generalization to theories with additional
gauge fields was considered in \cite{Ross94} and \cite{Emparan97},
whereas an extension to any $n\ge 4$ dimensions with arbitrary
dilaton coupling was given in \cite{GalRyt98}. In the special case
of Kaluza-Klein coupling, such a magnetizing transformation can be
interpreted as an appropriate dimensional reduction of a vacuum
$(n+1)$-dimensional spacetime \cite{Dowkeretal96}.

Here we review the case of $n$-dimensional pure Einstein-Maxwell gravity ($n\ge 4$) without any additional fields. The action is given by (from now on integrals are understood up to boundary terms)
\begin{equation}
 I=\frac{1}{16\pi}\int\d^nx\sqrt{-g}(R-F^2) ,
 \label{action}
\end{equation}
with $F^2=F^{\mu\nu}F_{\mu\nu}$ and $F_{\mu\nu}=A_{\nu,\mu}-A_{\mu,\nu}$.
Suppose we have a ``seed'' solution $(g_{\mu\nu},A_\mu)$ of the theory admitting a spacelike Killing vector $\pa_\phi$ with closed orbits such that, in adapted coordinates $\{x^i,\phi\}$, $i=1,\ldots,n-1$, one has $g_{i\phi}=0=A_i$. Explicitly, we assume
\beqn
 \d s^2 & = & \bar g_{ij}\d x^i\d x^j+V\d\phi^2 , \label{ansatzg} \\
 F & = & A_{\phi,i}\d x^i\wedge\d\phi , \label{ansatzF}
\eeqn
where $\bar g_{ij}\equiv g_{ij}$ represents the metric of a $(n-1)$-spacetime with coordinates $\{x^i\}$, $V\equiv g_{\phi\phi}$, and all the functions are independent of $\phi$. Then a new solution $(g'_{\mu\nu},A'_\mu)$ of the form (\ref{ansatzg}), (\ref{ansatzF}) (still admitting a Killing vector $\pa_\phi$) is generated by the transformation
\beqn
 \bar g'_{ij} & = & \Lambda^{2/(n-3)}\bar g_{ij} , \qquad V'=\Lambda^{-2}V , \nonumber \\
 A'_\phi & = & \Lambda^{-1}\left[A_\phi+B\left(\frac{1}{2}V+\frac{n-3}{n-2}A_\phi^2\right)\right] , \label{harrison} \\
 \Lambda & = & \left(1+\frac{n-3}{n-2}BA_\phi\right)^2+\frac{1}{2}\frac{n-3}{n-2}B^2V , \nonumber
\eeqn
where $B$ is a constant related to the strength of the transformed electromagnetic field (for $B=0$ Eq.~(\ref{harrison}) reduces to an identity transformation). We will consider always $B>0$. Following \cite{Dowkeretal94_49} (cf. also \cite{Emparan97}), the proof relies on showing that the action (\ref{action}) is invariant under the transformation (\ref{harrison}). First, it is convenient to use the above assumptions on the metric functions in order to reduce Eq.~(\ref{action}) to an effective $(n-1)$-dimensional action. The $n$-Ricci scalar $R$ can be decomposed as $R=\bar R-V^{-1/2}\bar g^{ij}(V^{-1/2}V_{,i})_{||j}$, where $\bar R$ and~$_{||}$ denote, respectively, the Ricci scalar and the covariant derivative associated with the $(n-1)$-metric $\bar g_{ij}$. Integrating over $\phi$ and using Eqs.~(\ref{ansatzg}) and (\ref{ansatzF}), the action (\ref{action}) becomes
\begin{equation}
 I=\frac{\Delta\phi}{16\pi}\int\d^{n-1}x\sqrt{-\bar g}V^{1/2}(\bar R-2V^{-1}\bar g^{ij}A_{\phi,i}A_{\phi,j}) .
 \label{reducedaction}
\end{equation}
Now we observe that the metric $\bar g_{ij}$ transforms
conformally under Eq.~(\ref{harrison}). Using the well known
relation between the Ricci scalars of conformal spaces and the
identity
$V^{1/2}(\ln\Lambda)_{||ij}=[V^{1/2}(\ln\Lambda)_{,i}]_{||j}-\frac{1}{2}V^{-1/2}V_{,j}(\ln\Lambda)_{,i}$,
one finds, by direct substitution of Eqs.~(\ref{harrison}) into
the action~(\ref{reducedaction}), that the latter is in fact
invariant.

\section{Alternative coordinates for extremal static rings}

\label{app_extreme}

The magnetized static rings (\ref{staticring}) become extremal when $\lambda=0$, in which case there is a regular, degenerate horizon at $y\to\infty$. Such extremal ``non-singular string loops'' were first constructed in \cite{Emparan01npb} using different coordinates (together with dilatonic solutions, which have a singular horizon). In this appendix we provide the explicit coordinate transformation between the two forms of the solutions. The metric of \cite{Emparan01npb} is
\beqn
 \d s^2= & & \Lambda\frac{\Delta+a^2\sin^2\theta}{\Sigma}\left[-\d t^2+r^2\cos^2\theta\d\psi_0^2+\frac{\Sigma^3}{[\Delta+(\tilde\mu+a^2)\sin^2\theta]^2}\left(\frac{\d 
    r^2}{\Delta}+\d\theta^2\right)\right] \nonumber \\
    & & {}+\Lambda^{-2}\left(\frac{\Sigma}{\Delta+a^2\sin^2\theta}\right)^2\Delta\sin^2\theta\d\phi_0^2 ,
\eeqn
where 
\beqn
 & & \Delta=r^2-a^2-\tilde\mu , \qquad \Sigma=r^2-a^2\cos^2\theta , \nonumber \\
 & & \Lambda=\left(1+\frac{1}{\sqrt{3}}B\frac{\tilde\mu a\sin^2\theta}{\Delta+a^2\sin^2\theta}\right)^2+\frac{1}{3}B^2
              \left(\frac{\Sigma}{\Delta+a^2\sin^2\theta}\right)^2\Delta\sin^2\theta ,
\eeqn
and $a$ and $\tilde\mu$ are constants. This expression turns out to be related to the line element (\ref{staticring}) [with $\lambda=0$, i.e. ${F(x)=1=F(y)}$] by the substitutions
\be
 \Delta=(1+\mu)^3L^2\frac{1-x}{x-y} , \quad \sin^2\theta=\frac{1+x}{x-y} , \quad \psi_0=\frac{1}{(1+\mu)^{3/2}}\,\psi , \quad  \phi_0=\frac{1}{(1+\mu)^{3/2}}\,\phi ,
\ee
along with the relation between the parameters
\be
 a^2=(1+\mu)^2(1-\mu)L^2 , \qquad \tilde\mu= 2\mu(1+\mu)^2L^2 .
\ee 
Analogously, the vector potential of \cite{Emparan01npb} transforms into our Eq.~(\ref{potential_ring}). Similar coordinates have been recently used to describe supersymmetric black rings in \cite{Elvangetal05}.


\providecommand{\href}[2]{#2}\begingroup\raggedright\endgroup

\end{document}